\documentclass[twocolumn, hyperref={colorlinks, linkcolor=blue, pdfborder={0 0 0}, breaklinks=true}]{aastex62}
\usepackage{graphicx}
\usepackage{xspace}
\usepackage{acronym}
\usepackage{color}
\usepackage{amsmath}
\usepackage{comment}
\usepackage[first=0,last=9]{lcg}

\usepackage{array,multirow}

\definecolor{LightCyan}{rgb}{0.88,1,1}

\usepackage{appendix}
\usepackage{tabularx}
\usepackage{longtable}
\usepackage{enumerate}
\usepackage{enumitem}
\usepackage{soul}
\usepackage{url}

\newcommand{\micmap}[1]{\textbf{\textcolor{blue}{MM: #1}}}

\newcommand{\abs}[1]{\lvert #1 \rvert}

\begin{document}

\title{\bf Search for Eccentric Black Hole Coalescences during the Third Observing Run of LIGO and Virgo}



\author{A.~G.~Abac}
\affiliation{Max Planck Institute for Gravitational Physics (Albert Einstein Institute), D-14476 Potsdam, Germany}
\author{R.~Abbott}
\affiliation{LIGO Laboratory, California Institute of Technology, Pasadena, CA 91125, USA}
\author{H.~Abe}
\affiliation{Graduate School of Science, Tokyo Institute of Technology, 2-12-1 Ookayama, Meguro-ku, Tokyo 152-8551, Japan}
\author{F.~Acernese}
\affiliation{Dipartimento di Farmacia, Universit\`a di Salerno, I-84084 Fisciano, Salerno, Italy}
\affiliation{INFN, Sezione di Napoli, I-80126 Napoli, Italy}
\author[0000-0002-8648-0767]{K.~Ackley}
\affiliation{University of Warwick, Coventry CV4 7AL, United Kingdom}
\author{C.~Adamcewicz}
\affiliation{OzGrav, School of Physics \& Astronomy, Monash University, Clayton 3800, Victoria, Australia}
\author{S.~Adhicary}
\affiliation{The Pennsylvania State University, University Park, PA 16802, USA}
\author[0000-0002-4559-8427]{N.~Adhikari}
\affiliation{University of Wisconsin-Milwaukee, Milwaukee, WI 53201, USA}
\author[0000-0002-5731-5076]{R.~X.~Adhikari}
\affiliation{LIGO Laboratory, California Institute of Technology, Pasadena, CA 91125, USA}
\author{V.~K.~Adkins}
\affiliation{Louisiana State University, Baton Rouge, LA 70803, USA}
% ...

\collaboration{The LIGO Scientific Collaboration, the Virgo Collaboration, and the KAGRA Collaboration}




\newcommand\hmm[1]{\ifnum\ifhmode\spacefactor\else2000\fi>1500 \uppercase{#1}\else#1\fi}

\newcommand{\GSTLAL}{GstLAL\xspace}
\newcommand{\CWB}{cWB\xspace}
\newcommand{\PYCBC}{PyCBC\xspace}
\newcommand{\MBTA}{MBTA\xspace}
\newcommand{\SPIIR}{SPIIR\xspace}
\newcommand{\MBTAONLINE}{MBTAOnline\xspace}
\newcommand{\LALSUITE}{LALSuite\xspace}
\newcommand{\LALINFERENCE}{LALInference\xspace}
\newcommand{\IDQ}{iDQ\xspace}
\newcommand{\BAYESWAVE}{BayesWave\xspace}
\newcommand{\BAYESTAR}{Bayestar\xspace}
\newcommand{\RIFT}{RIFT\xspace}
\newcommand{\PESUMMARY}{PESummary\xspace}
\newcommand{\GWCELERY}{GWCelery\xspace}

\newcommand{\BILBY}{{Bilby}\xspace}
\newcommand{\DYNESTY}{{Dynesty}\xspace}
\newcommand{\BILBYPIPE}{{BilbyPipe}\xspace}
\newcommand{\PBILBY}{{Parallel \BILBY{}}\xspace}
\newcommand{\ASIMOV}{{Asimov}\xspace}
\newcommand{\GWSUBTRACT}{{gwsubtract}\xspace}

\newcommand{\DMT}{{DMT}\xspace}
\newcommand{\DQR}{{DQR}\xspace}
\newcommand{\DQSEGDB}{{DQSEGDB}\xspace}
\newcommand{\GWDETCHAR}{{gwdetchar}\xspace}
\newcommand{\HVETO}{{hveto}\xspace}
\newcommand{\PYTHONVIRGOTOOLS}{{PythonVirgoTools}\xspace}
\newcommand{\OMICRONSCAN}{{Omicron}\xspace}

\newcommand{\HDF}{{HDF5}\xspace}
\newcommand{\PYTHON}{{Python}\xspace}
\newcommand{\NUMPY}{{NumPy}\xspace}
\newcommand{\SCIPY}{{SciPy}\xspace}
\newcommand{\PLT}{{Matplotlib}\xspace}
\newcommand{\SEABORN}{{seaborn}\xspace}
\newcommand{\GWPY}{{GWpy}\xspace}

\newcommand{\STTFOUR}{\texttt{SpinTaylorT4}}
\newcommand{\TFTWO}{\texttt{TaylorF2}}
\newcommand{\Phenom}{\texttt{Phenom}}
\newcommand{\IMRPhenomC}{\texttt{IMRPhenomC}}
\newcommand{\IMRPhenomD}{\texttt{IMRPhenomD}}
\newcommand{\IMRPhenomXHM}{\texttt{IMRPhenomXHM}}
\newcommand{\IMRPhenomXPHM}{\texttt{IMRPhenomXPHM}}
\newcommand{\SEOBNR}{\texttt{SEOBNR}}
\newcommand{\SEOBNRFOUR}{\texttt{SEOBNRv4}}
\newcommand{\SEOBNRP}{\texttt{SEOBNRv4P}}
\newcommand{\SEOBNRHM}{\texttt{SEOBNRv4HM}}
\newcommand{\SEOBNRPHM}{\texttt{SEOBNRv4PHM}}
\newcommand{\IMRPhenomPNRTidal}{\texttt{IMRPhenomPv2\_NRTidalv2}}
\newcommand{\IMRPhenomNSBH}{\texttt{IMRPhenomNSBH}}
\newcommand{\SEOBNRNSBH}{\texttt{SEOBNRv4\_ROM\_NRTidalv2\_NSBH}}
\newcommand{\SEOBNRROM}{\texttt{SEOBNRv4\_ROM}}
\newcommand{\NRSUR}{\texttt{NRSur7dq4}}

\newcommand{\CATONE}{Category~1\xspace}
\newcommand{\CATTWO}{Category~2\xspace}
\newcommand{\CATTHREE}{Category~3\xspace}

\newcommand{\LINSBUTRACTCHANNEL}{\reviewed{L1:LSC-POP\_A\_RF9\_I\_ERR\_DQ}\xspace}
\newcommand{\FLAGONETWOFOUR}{\reviewed{Flag~1.24 ($45~\mathrm{MHz}$ Sideband Fluctuations)}\xspace}

\newcommand{\FASTSCATTER}{\hmm{f}ast scattering\xspace}
\newcommand{\SLOWSCATTER}{\hmm{s}low scattering\xspace}
\newcommand{\BLIP}{\hmm{b}lip\xspace}
\newcommand{\TOMTE}{\hmm{t}omte\xspace}

\newcommand{\GWTCONE}{GWTC-1\xspace}
\newcommand{\GWTCTWO}{GWTC-2\xspace}
\newcommand{\GWTCTWOFINAL}{GWTC-2.1\xspace}
\newcommand{\GWTCTHREE}{GWTC-3\xspace}

\newcommand{\PLPEAK}{\textsc{Power Law + Peak}\xspace}

\acrodef{CWB}[\CWB{}]{coherent WaveBurst}

\begin{abstract}
Despite the growing number of confident binary black hole coalescences observed through gravitational waves so far, the astrophysical origin of these binaries remains uncertain. Orbital eccentricity is one of the clearest tracers of binary formation channels. Identifying binary eccentricity, however, remains challenging due to the limited availability of gravitational waveforms that include effects of eccentricity. Here, we present observational results for a waveform-independent search sensitive to eccentric black hole coalescences, covering the third observing run (O3) of the LIGO and Virgo detectors. We identified no new high-significance candidates beyond those that were already identified with searches focusing on quasi-circular binaries. We determine the sensitivity of our search to high-mass (total mass $M>70$\,$M_\odot$) binaries covering eccentricities up to 0.3 at 15\,Hz orbital frequency, and use this to compare model predictions to search results. Assuming all detections are indeed quasi-circular, for our fiducial population model, we place an upper limit for the merger rate density of high-mass binaries with eccentricities $0 < e \leq 0.3$ at $0.33$\,Gpc$^{-3}$ yr$^{-1}$ at 90\% confidence level.
\end{abstract}


\section{Introduction}
\label{sec:introduction}

The LIGO \citep{2015CQGra..32g4001L} and Virgo \citep{2015CQGra..32b4001A} gravitational wave observatories have completed three observing runs thus far. During these runs, 90 compact binary merger candidates were identified that had probability of astrophysical origin $p_{\rm astro}>0.5$ \citep{LIGOScientific:2021usb, 2021arXiv211103606T}. These discoveries opened previously inaccessible avenues to study the Universe, including the first direct information on binary black holes \citep{GW150914_Discovery, GW150914_Properties}, the multi-messenger observation of a binary neutron star coalescence \citep{2017ApJ...848L..12A, LIGOScientific:2017zic, Margutti:2020xbo}, a new type of constraint on cosmic expansion \citep{BNS_Hubble_2017, 2021arXiv211103604T}, and novel tests of general relativity \citep{LIGOScientific:2016lio, LIGOScientific:2017bnn, 2021arXiv211206861T}.

Despite the growing number of candidates and the insight they have  provided, the astrophysical sites and processes that produce the observed merging binaries remain uncertain. Multiple viable scenarios exist. The binary black holes could have formed in an isolated stellar binary \citep[e.g.,][]{1998ApJ...506..780B,2015ApJ...806..263D,2017MNRAS.468.5020I,2016A&A...588A..50M,2016MNRAS.460.3545D,2021ApJ...922..110G}, via dynamical interactions in dense stellar clusters (e.g., \citealt{2000ApJ...528L..17P,2010MNRAS.402..371B,2014MNRAS.441.3703Z,2015ApJ...800....9M,2016MNRAS.459.3432M,2016PhRvD..93h4029R,2017MNRAS.464L..36A}) or triple systems (e.g., \citealt{2017ApJ...841...77A,2020ApJ...903...67M,2021ApJ...907L..19V}), or via gas capture in the disks of active galactic nuclei (AGN; e.g., \citealt{2012MNRAS.425..460M,2017ApJ...835..165B,2019MNRAS.488...47F,2020ApJ...899...26T}). 

Gravitational waves carry information about the masses and spins of the merging black holes, which can be used to probe the binaries' origin \citep{GW150914_Properties, Vitale:2015tea, Zevin_2021}. Different formation channels have diverse predictions for the most common component masses, mass ratios, spin magnitudes and spin orientations \citep{Belczynski_2002, 2013ApJ...779...72D, Vitale:2015tea}. For example, isolated binaries are typically expected to produce black holes with spins mostly aligned with the binary's orbital axis with possible misalignments that could stem from  recoil velocities imparted during supernova explosion (e.g., \citealt{2016ApJ...832L...2R, 2018PhRvD..98h4036G,2019PhRvD.100d3012W}). 
Dynamically formed binaries, on the other hand, generally have an isotropic spin distribution (e.g., \citealt{2016ApJ...832L...2R,2017ApJ...840L..24F,2020PhRvD.102d3002B}). 
However, while masses and spins provide crucial information about the binaries' origin, there is often overlap between their distributions for various formation channels. A catalogue of binary black holes must therefore be considered to make statistical inferences about their origins using these properties alone.

Orbital eccentricity, $e$ is a unique signature that disfavors isolated binaries and favors triple systems, stellar clusters or AGN-assisted mergers as the possible formation scenario of the binary. While isolated black hole binaries can be born with an initial eccentricity, gravitational-wave emission will circularize their orbit by the time their orbital frequency reaches the sensitive band of ground-based gravitational-wave observatories \citep{Peters_1964}. Dynamical encounters can form binaries closer to merger, leaving insufficient time for orbital circularization. In AGN disks, eccentricity can be enhanced for a significant fraction of mergers, e.g., via binary--single interactions \citep{2022Natur.603..237S,2021ApJ...907L..20T}. Eccentricity can also be enhanced for field binaries by a nearby third object via the Kozai--Lidov mechanism \citep{1962_Kozai, 1962_Lidov, 2016ARA&A..54..441N,2017ApJ...841...77A,2018ApJ...853...93R,2023arXiv230210350B}. Identifying orbital eccentricity (or the lack thereof) in the population of binary black holes consequently places clear constraints on the proportion of binaries originating from various formation channels.

Despite the advantages that come with estimating the binary's orbital eccentricity, it has been difficult to probe this parameter through gravitational-wave observations for several reasons. (i) Eccentric orbits have wider dynamical range than quasi-circular, or $e=0$ orbits, making them more challenging to model semi-analytically \citep{PhysRevD.90.084016, PhysRevD.93.064031}. (ii) Eccentricity increases the dimension of the binary parameter space, requiring more gravitational waveform templates and substantially increasing the computational cost of both waveform computation \citep{Cornish:2010cd} and running template-based searches \citep{Lenon:2021zac}. (iii) Given these challenges and the lack of expected eccentricity in field binaries, the development of eccentric waveform models began with significant delay compared to circular waveform models \citep{1992MNRAS.254..146J}. Nonetheless, eccentric waveform development has been an active area recently, with several promising waveform models that can be useful in the future (e.g., \citealt{Hinderer:2017jcs,Cao:2017ndf,Liu:2021pkr,Nagar:2021gss, Albanesi:2021rby,Khalil:2021txt,Ramos-Buades:2021adz,Islam:2021mha,Setyawati:2021gom,Wang:2023ueg}).

While no comprehensive eccentric gravitational-wave template bank is currently available, indications of eccentricity already exist within the catalog of detected gravitational waves. The basis of such results is that standard gravitational-wave search algorithms developed to target circular binaries also have some sensitivity to eccentric binaries. For low masses $\lesssim 10$\,$M_\odot$, circular template-based searches show undiminished sensitivity for small residual eccentricities ($e\lesssim0.05$ at 40 Hz). To detect signals with eccentricities beyond $e \gtrsim 0.1$, we would however require template banks that include eccentric waveforms \citep{2010PhRvD..81b4007B}. In contrast, for higher masses and eccentricities, it has been shown that eccentricities can be found without significant loss of signal-to-noise ratio (SNR) using model-agnostic searches \citep{Abbott_2019_ebbh}.

To identify detected binaries as eccentric, two approaches have been carried out so far that circumvent the need for comprehensive template banks: 
\begin{itemize}
\item One approach is to employ Bayesian analyses using existing eccentric waveform models. An eccentric waveform model limited to eccentricities $e<0.2$ was used to show that the binary merger that produced the signal GW190521 as well as two others are consistent with originating from eccentric binary black holes (eBBH). \citep{2020ApJ...903L...5R,2021ApJ...921L..31R}. Using a different waveform model that includes the full eccentricity range, \cite{2021arXiv210605575G} found strong support for the binary coalescence that produced GW190521 being highly eccentric. Both models were limited to waveforms with black hole spins aligned with the binary orbit. Orbital eccentricity and misaligned spins that induce precession of the orbital plane produce similar imprints in the gravitational wave, and both of these effects should preferably be accounted for in order to accurately analyze the event \citep{CalderonBustillo:2020xms, Romero-Shaw:2022fbf}. 
\item A different approach relies on numerical relativity simulations of eBBHs. Due to the computational cost, only a limited number of simulations can be carried out, which can only sparsely cover the parameter space. \cite{GayathrieBBH} used such numerical relativity waveforms that discretely cover the full eccentricity space and includes waveforms with both aligned and misaligned spin with the binary orbit. Interpolation methods and consistency checks were applied to recover the eccentricity and other parameters of the binary. They found that the signal GW190521 is most consistent with being produced by a highly eccentric ($e\sim0.7$) binary.
\end{itemize}
 The GW190521 signal for which the above analyses were applied was already considered special even without the indication of eccentricity, having had a high reconstructed total black hole mass of $153.1^{+42.2}_{-16.2}$\,$M_{\odot}$, along with high and probably misaligned spin \citep{gw190521_implications}.

In this paper, we carry out a search focusing on eccentric black hole coalescences over the third observing run (O3) of the LIGO--Virgo network. We use a minimally modeled search algorithm \citep{Klimenko:2005xv,Salemi:2019uea,2016PhRvD..93d3007T} that we optimize for sensitivity for a set of high-mass (total mass $M\geq70$\,$M_\odot$), eccentric gravitational waveforms \citep{Hinder:2017sxy, SXS_Collab_2019}. As methods to estimate the eccentricity of individual events are under development, we instead focus on potential detections that have not already been discovered by other searches, and characterize the sensitivity of our search to eccentric binaries, relying on methods with well understood performance.

The paper is organized as follows. In Section \ref{sec:algorithm} we introduce our search algorithm and demonstrate its sensitivity to eccentric waveforms. In Section \ref{sec:results} we present our search results. In Section \ref{sec:pop} we discuss constraints on astrophysical populations based on our search results. We conclude in Section \ref{sec:conclusion}.

Gravitational wave strain data \citep{gwosc:gwtc3} and posterior samples \citep{ligo_scientific_collaboration_and_virgo_2021_5546663} for all events from GWTC-3 are available from the Zenodo platform or the Gravitational Wave Open Science Center \citep{Abbott:2019ebz}.

\section{Search algorithm and sensitivity}
\label{sec:algorithm}

\subsection{Characterization of eccentricity}
\label{sec:ecc}

Due to the emission of gravitational waves, binary orbits have a gradually decreasing orbital separation. Eccentric binary orbits also circularize over time due to the emission of gravitational waves \citep{Peters_1964}. This makes the definition of eccentricity challenging. Determining eccentricity is particularly difficult at the late stages of the binary evolution when less than a full orbit separates the black holes from merger. 

There have been various efforts to define eccentricity for binary compact object systems. These  eccentricity definitions involve Keplerian orbit assumptions \citep{1963PhRv..131..435P,Loutrel_2018},  angular frequencies at apocenter and pericenter  \citep{Mora_eccentricity}, calculations using instantaneous radial acceleration \citep{PhysRevD.97.064027} and using coordinate separations \citep{coord_sep_ecc}. A detailed list of the different eccentricity definitions that have been developed so far can be found in \cite{Loutrel_2018}.

For our analysis, we adopt the eccentricity definition following \cite{2022PhRvD.106l4040R}, based on calculation first developed by \cite{Mora_eccentricity} and later used by \cite{Lewis_2017}, \cite{ Ramos-Buades:2019uvh} and \cite{ Shaikh:2023ypz}. To compute eccentricity for each orbit, we used the gravitational-wave frequencies at apocenter ($\omega_a$) and the consecutive pericenter ($\omega_p$).  With these, eccentricity for the given orbit is 
\begin{equation} \label{eq:2a}
    e = \cos(\psi/3) - \sqrt{3} \sin(\psi/3)
\end{equation}
with
\begin{equation} \label{eq:2b}
    \psi = \arctan \left( \frac{1 - e_{22}^2}{2 e_{22}} \right),
\end{equation}
where 
\begin{equation} \label{eq:1}
    e_{22} = \frac{\sqrt{\omega_p} - \sqrt{\omega_a}}{\sqrt{\omega_p} + \sqrt{\omega_a}}.
\end{equation}
We used the orbital frequency of the $\ell=2$, $m=2$ multipole moments of the gravitational-wave signal. 

In order to characterize the eccentricity as a function of time, we associate this eccentricity with a frequency that is an average of the pericenter and apocenter frequencies. This method of computing eccentricity using the waveform itself is advantageous because (i) it enables us to compute the evolution of eccentricity as a function of time (and frequency); (ii) it is gauge independent; and (iii) this definition can be uniformly applied to all waveform models and can be computed during post-processing. 
We quote eccentricity values at 15\,Hz gravitational-wave emission frequency unless specified otherwise. We choose this specific value as this is approximately the low-frequency limit of LIGO--Virgo network's sensitivity, and is therefore of the order of the initial frequency of detected gravitational-wave signals. This also compares well to the frequency at which eccentricity is typically quoted by different astrophysical models (usually defined at a gravitational-wave emission frequency of $\sim 10-15$\,Hz ; e.g., \citealt{Fragione:2019dtr, 2021ApJ...921L..43Z}).

\subsection{Eccentric waveforms}

There are multiple ongoing efforts to develop a comprehensive set of eccentric binary coalescence waveforms. Multiple waveform families have been generated using the semi-analytical effective-one-body formalism, which are currently restricted to non-precessing spins \citep{Nagar:2021gss,Ramos-Buades:2021adz}. A suite of numerical relativity simulations have also been carried out that cover virtually the full eccentric and spin parameter space \citep{GayathrieBBH,2022arXiv220200018H}. 

For our analysis, we adopted 12 state-of-the-art numerical relativity waveforms from the Simulating eXtreme Spacetimes (SXS) Collaboration \citep{Hinder:2017sxy, SXS_Collab_2019}, which were the only high-fidelity waveforms available to us at the time of this study. These waveforms cover the eccentricity space up to 0.3 defined at 15\,Hz gravitational-wave frequency, and include a range of mass ratios: $q\equiv m_2/m_1 = \{1,0.5,0.33\}$, where $m_2$ and $m_1$ are the lighter and heavier masses, respectively. 

\begin{figure*}
\centering
\includegraphics[scale=0.55]{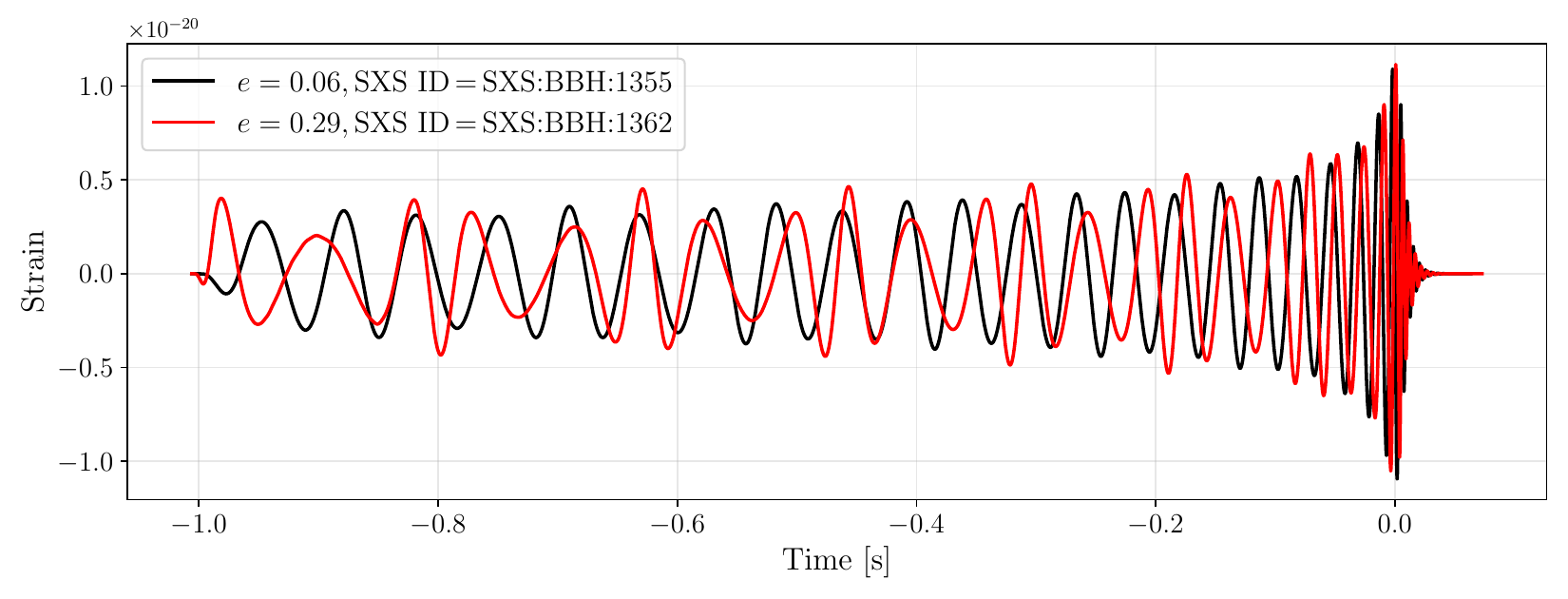}
\caption{Examples of time-domain waveforms with two different eccentricities (indicated in the legend) for equal mass binary systems with total source mass of $90 \, M_{\odot}$ at a distance of $100$ Mpc. The simulations start at an orbital separation that translates to an orbital frequency $f_\mathrm{low}=15\,\mbox{Hz}$. The eccentricity values indicated in the legend are defined at the same $f_\mathrm{low}$.}
\label{fig:waveform_example}
\end{figure*}

As the numerical relativity simulations were carried out for the late stage of the binary coalescence, they cover the gravitational waveform for the full frequency band of the ground-based detectors only for total binary source masses $\gtrsim 70$\,$M_\odot$. Above this mass limit any binary mass can be obtained by a simple scaling of the simulated waveforms due to the scale invariance of general relativity \citep{Tiglio:2019yec}. The selected waveforms are non-spinning, which has limited effect on the sensitivity estimates we compute below. When reconstructing the properties of detected gravitational-wave signals, it is important to include spins, as eccentricity and spin precession can mimic each other \citep{CalderonBustillo:2020xms, Romero-Shaw:2022fbf}. Since we do not use these waveforms to reconstruct properties of signals in this analysis, this problem is not relevant here. We list the properties of the waveforms in Table \ref{tab:NR}. Figure \ref{fig:waveform_example} shows the change in signal morphology as the orbital eccentricity is changed while keeping other source parameters fixed.

\begin{table}
\begin{center}
\begin{tabular}{ c  c  c }
\hline
 $q$ & $e$ & Waveform ID\\
\hline
 $0.33$ & $0.08$ & SXS:BBH:1371\\
 
 $0.33$ & $0.12$& SXS:BBH:1372\\

 $0.33$ & $0.27$ & SXS:BBH:1374\\

 $0.5$ & $0.09$ & SXS:BBH:1365\\

 $0.5$ & $0.14$ & SXS:BBH:1366\\

 $0.5$ & $0.29$ & SXS:BBH:1369\\

 $0.5$ & $0.30$ & SXS:BBH:1370\\

 $1.0$ & $0.06$ & SXS:BBH:1355\\

 $1.0$ & $0.14$ & SXS:BBH:1357\\

 $1.0$ & $0.22$ & SXS:BBH:1361\\

 $1.0$ & $0.29$ & SXS:BBH:1362\\

 $1.0$ & $0.30$ & SXS:BBH:1363\\
 
\hline 
\end{tabular}
\caption{Parameters of the 12 numerical relativity simulations adopted from the SXS binary black hole simulations catalog \protect{\citep{SXS_Collab_2019}}. Columns show the binary's mass ratio $q$, and eccentricity $e$ at a reference emission frequency of $15 \, \rm{Hz}$ (Section \ref{sec:ecc}) for a binary source total mass of $90 M_{\odot}$. Spin amplitudes $\chi_1$ and $\chi_2$ are zero for all considered models.}
\label{tab:NR}
\end{center}
\end{table}

We used this set of 12 numerical relativity waveforms to quantify the search sensitivity to high-mass ($\gtrsim 70$\,$M_\odot$) eccentric black hole mergers. However, with this limited set of waveforms we could not reconstruct the eccentricity of events. 

\subsection{Search optimization and sensitivity improvement} 
\label{sec:optimization}

Current template-based searches \citep{CANNON2021100680, Aubin:2020goo, Nitz:2017svb} do not include eccentric gravitational waveforms. As a consequence, their sensitivity is limited for such events, in particular at high eccentricities and low masses \citep{2010PhRvD..81b4007B}. Our search was therefore based on the coherent WaveBurst algorithm (cWB; \citealt{Klimenko:2005xv,2016PhRvD..93d3007T,Salemi:2019uea}), which uses minimal assumptions about the signal waveform and hence is expected to be sensitive to eccentric signals. 

The cWB algorithm uses the Wilson--Daubechies--Meyer filter to transform time domain detector data to time--frequency representations \citep{cWB_WDM}. Excess power regions in the time--frequency representation of strain data that are obtained from the network of detectors are then identified by cWB using clustering algorithms. Selected clusters with excess energy above the expected detector noise are identified as events. The signal waveform, sky coordinates and waveform polarization of the source are then reconstructed for these events using maximum-likelihood analysis \citep{cWB_MaxL}. 

Once the search pipeline is run, thresholds are placed by cWB on the coherent statistics that it derives for each candidate event. These are used to better differentiate between astrophysical signals and noise artifacts \citep{gayathri_wavegraph}. We will refer to these thresholds on cWB statistics as vetoes. Vetoes define a part of the parameter space over the coherent statistics that should be excluded from the analysis due to the high rate of non-Gaussian noise artifacts there. To maximize the sensitivity of cWB to eccentric binaries, we carried out an optimization of these vetoes applied by cWB to each event. The first two sets of vetoes that are common to the standard cWB pipeline and the eccentric search pipeline are summarized in Appendix \ref{sec:vetoes}.

Transient non-Gaussian noise artifacts, also known as \textit{glitches}, can limit the detector's sensitivity to gravitational-wave signals. Targeted vetoes are placed by the standard cWB pipeline to mitigate this problem. These glitch-focused vetoes are derived using cWB summary statistics $Q_a$ and $TF$. The waveform shape parameter derived by cWB is denoted by $Q_a$, and is a function of another cWB parameter Qveto ($Q_a = \sqrt{\mathrm{Qveto}}$). This parameter quantifies how well the total energy of the signal is distributed across time \citep{Qveto,gayathri_wavegraph,Mishra:2021tmu}. The threshold $Q_a>0.3$ is placed to better distinguish between gravitational waves and a class of low-frequency transient noise artifacts called Blip glitches \citep{Cabero_2019, Davis_2021}. Signals due to Blip glitches, which have most of their energy localized to a small time segment have low $Q_a$ values as opposed to signals from binary coalescence, which have higher $Q_a$ values as a consequence of signal energy being distributed over a longer duration. The $TF$ parameter is a function of the signal bandwidth, duration, and power which are additional statistics that cWB estimates for candidate events. A threshold on this parameter is placed to ensure that short-duration glitches that mimic gravitational-wave signals from intermediate mass binary black hole systems are removed. 

\begin{figure}
\center
\includegraphics[scale=0.6]{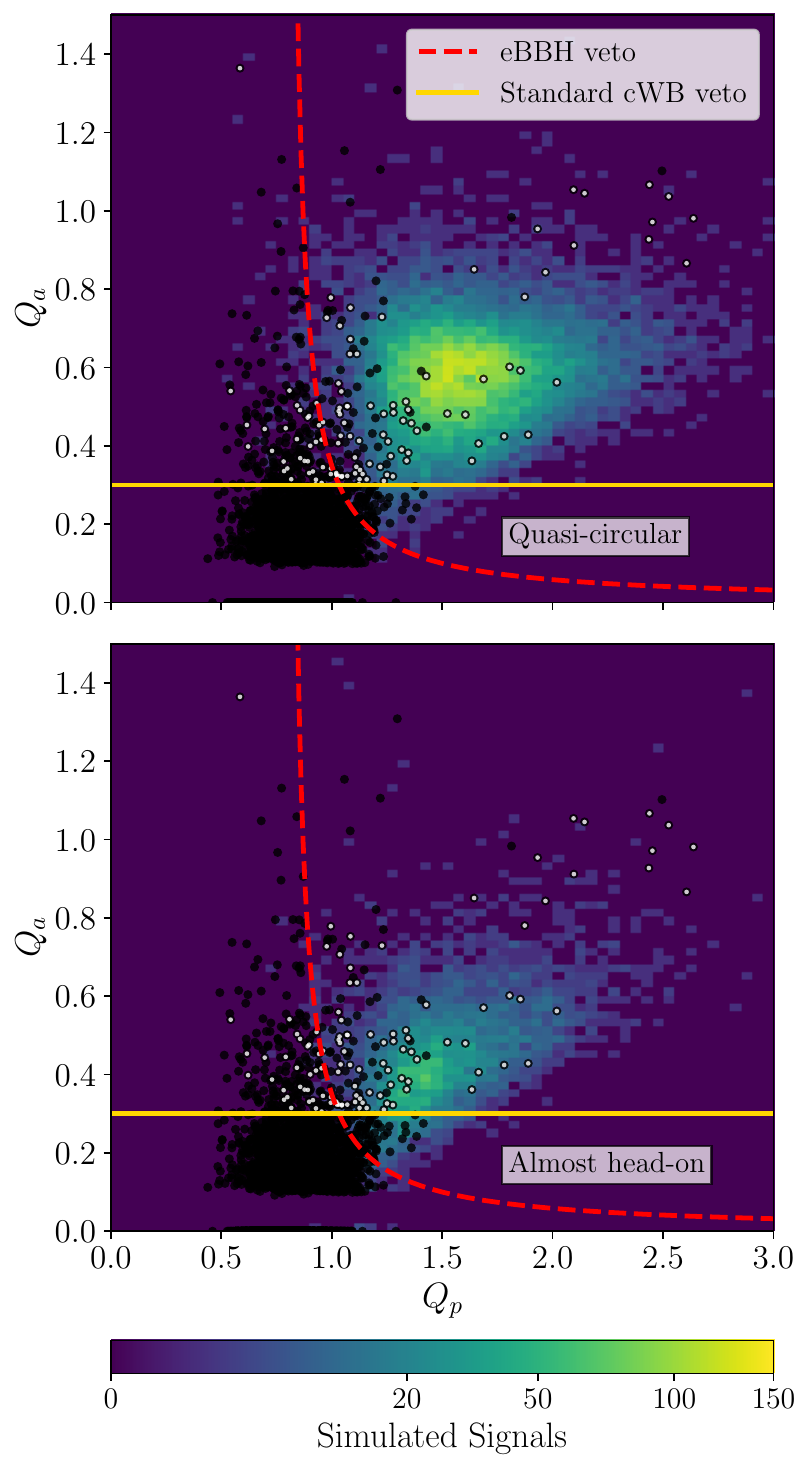}
\caption{Distribution of $Q_a$ and $Q_p$ for simulated signals (shown as two-dimensional histogram with colorbar denoting number of events in each two-dimensional bin) and loud simulated background events (shown as black dots). The cWB statistics $Q_a$ and $Q_p$ describe the morphology of a signal. The yellow line represents the standard cWB $Q_a$ veto and the red dashed line denotes the eBBH $Q_{a}$--$Q_{p}$ veto. The white dots correspond to loud background events that remain after all standard cWB vetoes \citep{Lopez:2021ikt} are applied. \textit{Top}: Simulated signals correspond to equal mass, source total mass, $M=150 \, M_{\odot}$, quasi-circular orbit systems. \textit{Bottom}: Simulated signals correspond to equal mass, $M=150 M_{\odot}$, almost head-on (highly eccentric) systems.}
\label{fig:Qa_Qp}
\end{figure}

We injected simulated gravitational-wave signals from equal mass, almost head-on systems \citep{2022arXiv220200018H} into real detector data to find the set of vetoes that do not remove highly eccentric signals while still rejecting most noise artifacts. To perform this optimization, the cWB algorithm was used to detect these injected signals and derive their properties. Vetoes were selected such that they maximized the number of detections at fixed false alarm rates. 

We observed that $Q_a$ and $TF$ vetoes were prone to removing a significant fraction of highly eccentric simulated signals. We found that we could mitigate this problem if we removed these two thresholds, and instead introduced a new $Q_a$--$Q_p$ veto to better distinguish between signals from highly eccentric binaries and short-duration glitches. This veto removes events identified by cWB that do not satisfy the condition  $Q_{a}(Q_{p}-0.8)>0.07$. The summary statistic $Q_p$ quantifies the number of cycles in the reconstructed signal. The $Q_{a}$--$Q_{p}$ veto along with the first two sets of vetoes from the standard search which are summarized in Appendix \ref{sec:vetoes} were selected as the set of post-production vetoes for the eBBH search. We will refer to this version of cWB that is optimized to eccentric mergers as cWB-eBBH. While the vetoes were optimized using equal-mass waveforms, we confirmed that the optimized search improved eccentric event recovery for unequal mass injections as well.

Figure \ref{fig:Qa_Qp} shows an example of the standard-cWB $Q_a$ veto and the new cWB-eBBH $Q_a$--$Q_p$ veto for quasi-circular and highly eccentric systems. We also look at this veto's performance with background events. To generate background events, data from one detector is time-shifted relative to the other detector's data by an amount greater than the maximum time for a gravitational wave signal to travel between the detectors \citep{LIGOScientific:2016gtq}. The standard veto does well in removing background events and recovering the majority of quasi-circular simulation events. However, the distribution of simulation signals in the $Q_a-Q_p$ space changes for highly eccentric systems and as a consequence, the standard cWB veto removes a significant fraction of simulation events.

We characterize the sensitivity improvement due to the optimization procedure by computing the number of injected gravitational waves detected by cWB-eBBH but not by standard cWB, divided by the total number of detections by standard cWB. Here we consider a signal detected if it corresponds to an inverse false alarm rate (IFAR) of $\geq 1~\mathrm{yr}$. This IFAR threshold of $\geq 1~\mathrm{yr}$ was only used to assess the improvement in sensitivity from the introduction of the cWB-eBBH veto, and not as a general detection threshold.

The fraction of events recovered with IFAR $\geq 1~\mathrm{yr}$ by cWB-eBBH that are removed by the standard pipeline with respect to the total number of events recovered by the standard pipeline is $\sim 28 \%$ for head-on collision (highly eccentric) equal mass systems with a source total mass of $150 \, M_{\odot}$. Additionally, we see that this fraction is higher ($\sim 34 \%$) for systems with more unequal mass. Therefore, our optimization is the most significant for highly eccentric binaries with unequal masses. The performance of cWB-eBBH for low eccentricity signals remains comparable (within 5\%) to the standard pipeline. We conclude that the cWB-eBBH veto does significantly better than the standard veto to improve sensitivity for highly eccentric systems without degrading sensitivity to less eccentric systems.

\section{Results}
\label{sec:results}

\subsection{Search sensitivity}
\label{sec:sensitivity}

We carried out a search for simulated gravitational-wave signals to quantify the sensitivity of the cWB-eBBH search algorithm. We performed injections in offline (high-latency) re-calibrated O3 strain data with category 0, 1, 2 and 4 data-quality vetoes \citep{Davis_2021, 2021arXiv211103606T}. Category~0 vetoes are applied to ensure that the segments of data used in this analysis were collected when the detectors were in observing mode. Category~1 vetoes are used to discard data from periods in which the detectors were running in an improper configuration, data-dropout or on-site maintenance occurred at either detector, or when there are major problems with the operation of an instrument at the detectors. Category~2 vetoes flag data segments that likely contain non-Gaussian noise artifacts. Category~4 vetoes flag data segments that contain hardware injections. 
The injected waveforms have source total mass $M \in [70 \, M_{\odot}, 200 \, M_{\odot}]$. We used the possible 12 configurations of $e$ and $q$, with 6 choices of source total mass for each of these configurations. Waveforms with different masses were obtained by scaling each of the 12 numerical relativity waveforms listed in Table \ref{tab:NR}. The simulated signals for each fixed set of source parameters of $(M, e, q)$ were uniformly distributed in sky location ($\theta, \phi$) and inclination $\iota$. They were also distributed uniformly in co-moving volume up to a maximum redshift $z_\mathrm{max}$. For each waveform, we separately calculated $z_\mathrm{max}$ up to which they must be injected so that we do not make unnecessary injections that the search cannot detect. This was calculated with an optimal two-detector-network (Livingston--Hanford) signal-to-noise-ratio threshold of 5.0. Since we observe signals with redshifted mass \citep{Schutz_Redshift}, it is in principle possible to inject simulations with total source mass $< 70$\,$M_\odot$ if we populate them at higher redshifts. This was however not performed in the presented analysis. Injections spaced uniformly in time approximately every 100\,s in the O3 dataset. 

We used the fraction of detected and injected waveforms to compute the sensitive distance of the search for the given waveform. Sensitive distance \citep{2019PhRvD.100f4064A} is defined such that a detector that detects every event within the sensitive distance and no event beyond, it would have the same detection rate as our detector network.

A similar analysis was carried out with data from the first two observing runs of LIGO--Virgo using approximate eccentric waveform models \citep{Abbott_2019_ebbh}. This analysis spanned the binary mass parameter space from $10 \, M_{\odot}$ to $100 \, M_{\odot}$ while the analysis described in this paper covers binary mass of $70 \, M_{\odot}$ to $200 \, M_{\odot}$. The sensitivities reported in this paper are higher than that analysis due to increased sensitivity of the detector during the third observing run, and due to the higher masses considered here. There have also been studies to characterize the effect of eccentricity in the sensitivity of long-duration signals with unmodeled search pipelines using hybrid inspiral--merger--ringdown waveform models \citep{LD_eBBH_O3}. However, these studies were targeted towards low mass binary black holes and binary neutron stars as opposed to our search, which is targeted towards high mass eccentric binaries. Therefore, the search sensitivities reported in \cite{LD_eBBH_O3} are lower than what we obtain in this paper.

\begin{figure}
\center
\includegraphics[width=0.48\textwidth]{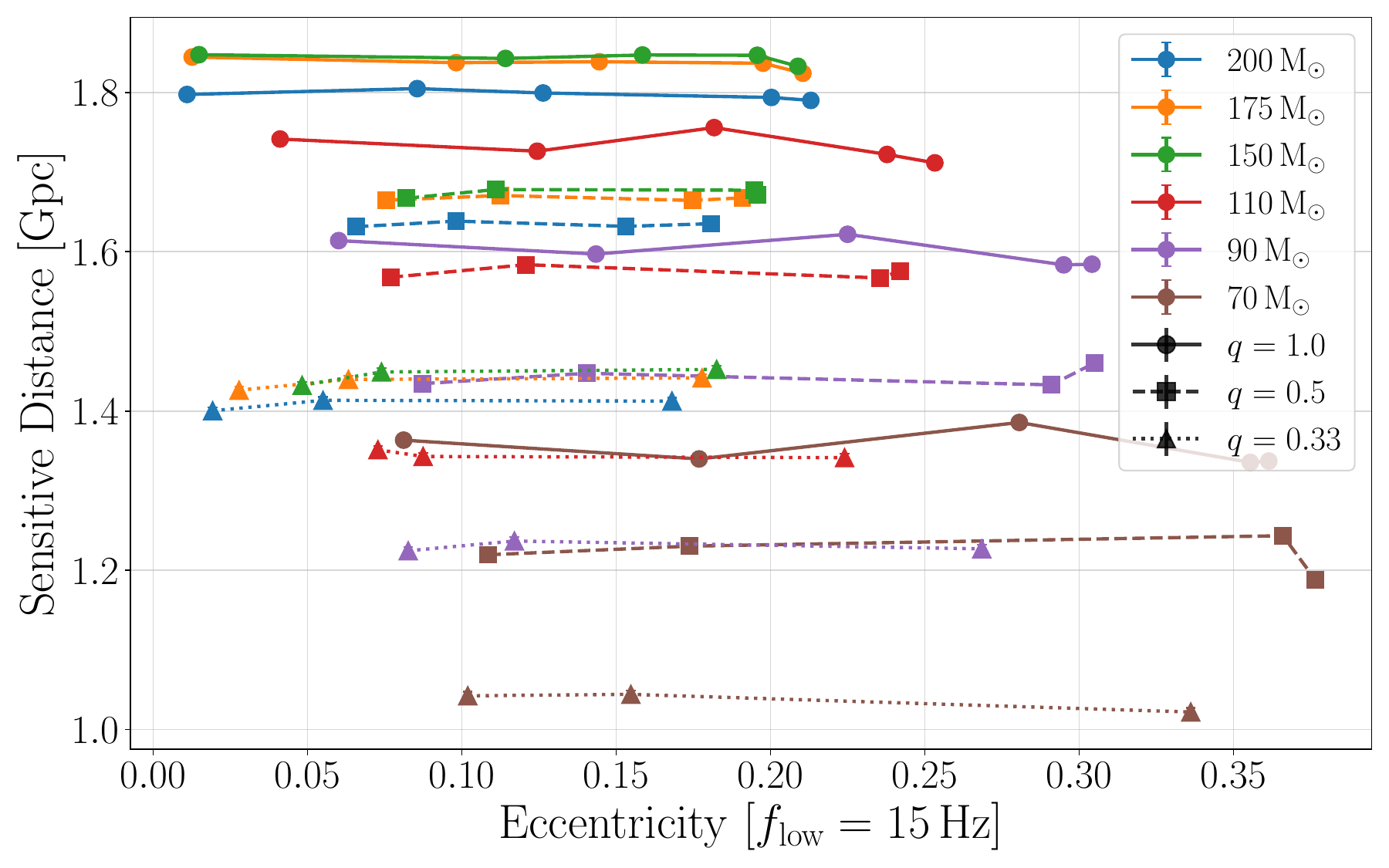}
\caption{Sensitive distance as a function of orbital eccentricity for different binary total masses and mass ratios. Different marker shapes represent systems with different mass ratios and the different colors represent the various total masses
considered here. We used an IFAR threshold of $1.32$\,yr, which was the loudest new candidate's IFAR. The horizontal axis denotes the eccentricity of the binary at an orbital separation that corresponds to a frequency of $15 \, \rm{Hz}$. The statistical error bars on the obtained sensitive distance are smaller than can be presented on this plot.}
\label{fig:sensitive_distance_figure}
\end{figure}

The obtained sensitive distance is shown in Figure~\ref{fig:sensitive_distance_figure}, for different source total masses and mass ratios, as a function of binary eccentricity. The statistical error bars for the obtained sensitive distance range between 0.21\,Mpc and 5.63\,Mpc. We see that the sensitivity at the considered high masses is mostly independent of the eccentricity up to our highest eccentricity of 0.3. We also see, as expected, that sensitivity is highest for equal mass binaries, and gradually drops as the difference between the two black hole masses increases.

\subsection{Search and loudest event}

We carried out the cWB-eBBH search over the third observing run of the LIGO and Virgo detectors. For most of the observing run we used data from only the two LIGO detectors, as search sensitivity was not appreciably affected by the addition of Virgo data. For the January 4, 2020 to January 22, 2020 period we also incorporated Virgo in the search to analyze the candidate 200114\_020818, which was found by the intermediate mass black hole binary search \citep{O3_IMBH} in the three detector network configuration comprising the LIGO and Virgo detectors. Follow-up studies for this event \citep[][Appendix B]{O3_IMBH} showed inconsistent results under a quasi-circular binary black hole hypothesis. We investigated if this candidate had higher significance under the eccentric hypothesis. However, this candidate was removed by the cWB-eBBH vetoes. The search and sensitivity results presented below were obtained using data from only the two LIGO detectors.

\begin{figure}
\center
\includegraphics[width=0.47\textwidth]{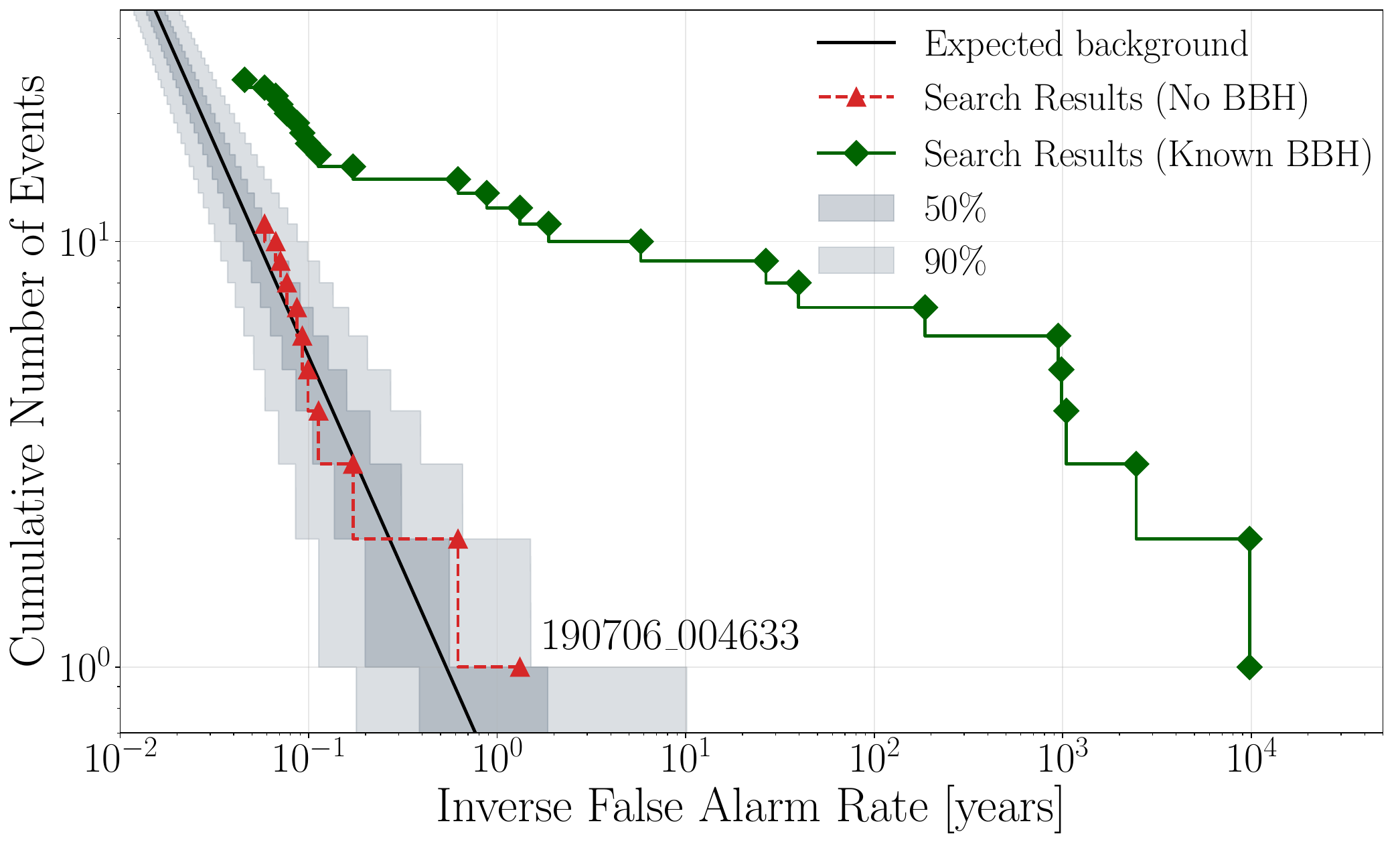}
\caption{Cumulative number of events as a function of IFAR recovered by the cWB-eBBH search. The solid line represents the expected background for the O3 search, and the gray regions correspond to the 50\% and 90\% Poisson uncertainty regions. Green squares denote previously reported gravitional-wave candidates \citep{2021arXiv211103606T, O3_IMBH} recovered by our search, and red triangles show events that were not previously reported by other searches.}
\label{fig:open_box_plot}
\end{figure}

Our search recovered 28 gravitational-wave candidates with IFAR\,$>1$\,yr. By choosing this IFAR threshold, we eliminate low significance candidates that could have been due to noise artifacts in the detector. All but one of these events have been identified previously by other searches as well \citep{2021arXiv211103606T, O3_IMBH}. The results of our search are summarized in Figure \ref{fig:open_box_plot}. The search results excluding previously found candidates is consistent with background noise. 

\begin{figure*}
\centering
\includegraphics[scale=0.52]{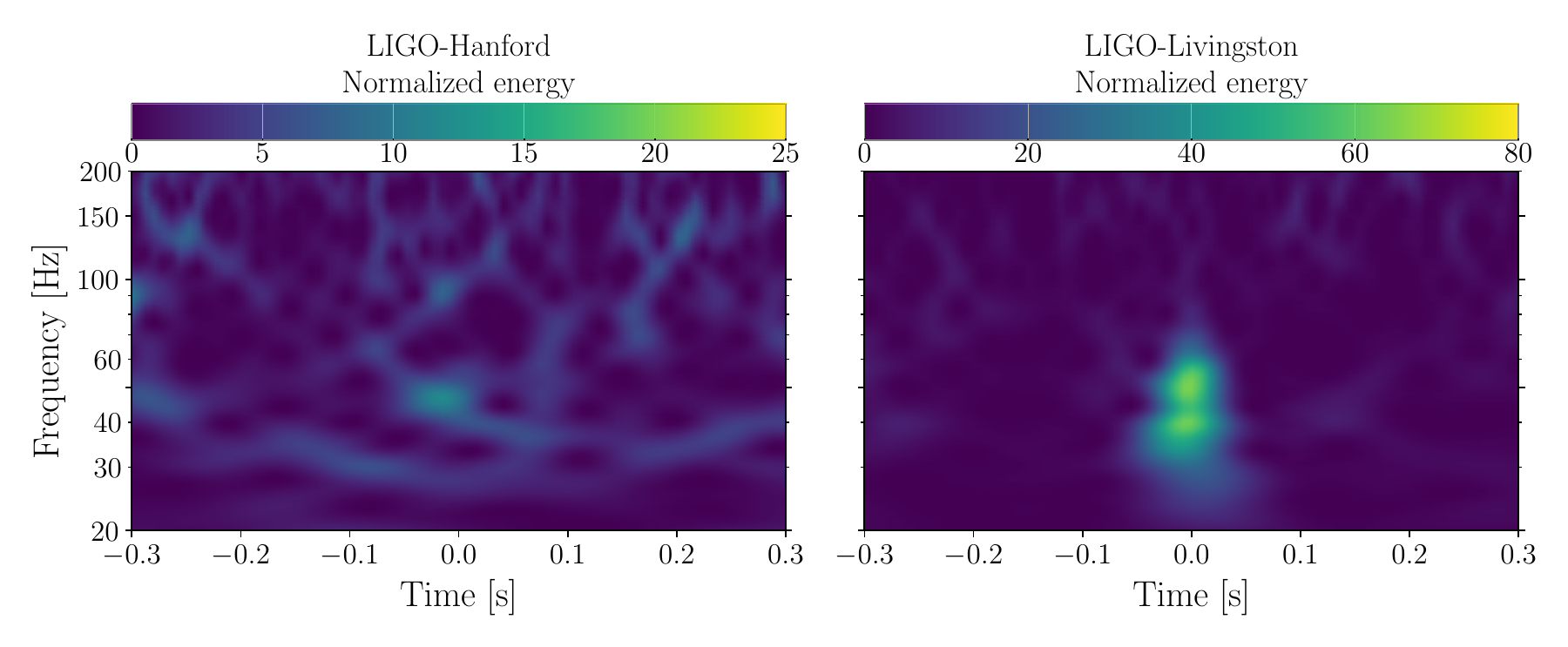}
\caption{Time-frequency map (spectrogram) of the most significant new candidate identified by the cWB-eBBH search. We show the spectrogram for the LIGO-Hanford (left) and LIGO-Livingston (right) detectors. The individual detector SNRs in the LIGO-Hanford and LIGO-Livingston are 5.6 and 10.9 respectively. Since the energies in the two detectors are very different, we use different scales on the colorbar. The Virgo detector was in observing mode during the time of this event. We used data from all three detectors for follow-up studies and observed that the SNR in the Virgo detector for this event was low ($\sim 2$).}
\label{fig:event_figure}
\end{figure*}

We identified one event candidate with an IFAR\,$>1$\,yr that was not previously reported. This most significant new candidate, hereafter referred to as 190706\_004633, was observed on July 6, 2019. It was recovered with an IFAR of $1.32$\,yr. It has an SNR of $12.2$ and a central frequency of $74\,$Hz. Figure \ref{fig:event_figure} shows the time--frequency map of this event candidate. 

In order to better understand whether 190706\_004633 is of astrophysical origin, we carried out a detailed study of the detector performance and characteristics at the time of the event. This study was aimed to uncover signs of instrumental or environmental artifacts that could have altered the gravitational wave data and hence produced the candidate \citep[][Section 3.2.4]{Davis_2021}. No such artifacts were found. However, the Gravity Spy machine learning classifier \citep{Zevin:2016qwy, Soni:2021cjy} classified the excess power in LIGO Livingston as a Tomte glitch. Tomtes are a common glitch class that are similar in morphology to high-mass binary coalescence signals \citep{Ashton:2021tvz}. No glitch or signal was identified in the LIGO Hanford data by the same classifier. However, as the Gravity Spy machine learning model is not designed to search for astrophysical signals \citep{Glanzer:2022avx} or to differentiate eccentric binary black hole merger signals from glitches, we cannot rule out an astrophysical origin. 

To further investigate this event we carried out a standard parameter estimation analysis of the data using LALInference \citep{2015PhRvD..91d2003V} with nested sampling assuming a quasi-circular waveform. We investigated properties of this event using data from the two LIGO detectors as well as the Virgo detector. 
For this analysis, in lieu of an eccentric waveform that fully covers the necessary parameter space, we adopted the quasi-circular binary approximant IMRPhenomXPHM \citep{Pratten:2020ceb}. This estimation found that the estimated source total mass of 190706\_004633 is $M\sim 320\, M_{\odot}$, and its estimated redshift is $z\sim0.3$. Studies have shown that the chirp mass of a binary with low to moderate eccentricity can be reconstructed with a bias of up to 4\,\% using parameter estimation with quasi-circular waveforms \citep{OShea:2021ugg}. However, the reconstructed parameters would be considerably more inaccurate if the signal originated from a highly eccentric binary. Therefore, these results indicate that the signal, if astrophysical, would correspond to a high-mass binary, but should not be used to give precise indications of source properties. 

Although astrophysical origin could not be ruled out, we conclude from the large difference measured in the LIGO Hanford and Livingston SNRs that this event is in accordance with an incoherent noise origin rather than a binary black hole origin. In the following section we therefore compute upper limits to merger rates assuming non-detection of any eccentric event.

\section{Eccentric Binary Population Models}
\label{sec:pop}

In order to understand the astrophysical implications of our results, we computed the expected number of detections for a fiducial source model. For this we adopt the joint total mass and mass ratio probability density $p(M,q)$ which was found to be the best fit for LIGO--Virgo's observations listed in the GWTC-3 catalog \citep{2021arXiv211103606T, KAGRA:2021duu} assuming the Power Law + Peak model described in \cite{GTWC2}. As we have waveforms and simulations that are sparsely sampled in mass and mass ratio, we linearly interpolated the sensitivity of the existing waveforms to points in between the available points in order to obtain a sensitive distance for any source total mass and mass ratio within $70\,M_\odot\leq M\leq 200\,M_\odot$ and $0.33<q<1.0$. For a more general distribution, we considered a power-law black hole mass distribution of $M^{-2.3}$ (assuming a Salpeter initial mass function; \citealt{Perna:2019axr}) and a uniform distribution in mass ratio. We further adopted an eccentricity distribution in which the probability density of the binaries' eccentricity is $p(e) \propto 2(1-e)$. This  distribution is chosen to characterize a population which has a larger fraction of low eccentric binaries.

Having defined the probability density of our fiducial population with respect to the binary parameters, using the sensitive distance obtained over the considered parameter space (see Section \ref{sec:sensitivity}), we computed the total volume--time $\mbox{VT}$ \citep[][Appendix A]{2019PhRvD.100f4064A} covered by our search during O3, assuming an IFAR threshold of 1.32\,yr, which is the IFAR of our search's loudest new event. For our fiducial model, we obtained $\mbox{VT}=$ 6.88 Gpc$^{3}$\,yr for eccentric binaries with $0<e<0.3$. 
Assuming non-detection of any eccentric event, this would correspond to a constraint of $<0.33$\,Gpc$^{-3}$ yr$^{-1}$ on the merger rate density at 90\% confidence level in the $70\,M_\odot\leq M\leq 200\,M_\odot$ and $0.33<q<1.0$ parameter space. 


With the small number of available eccentric waveforms for this study, we cannot determine if discovered binaries are eccentric. Therefore, we cannot discount the possibility that previously identified gravitational-wave candidates originate from eccentric binaries. In this case, the number of observed eccentric binaries is greater than zero, and so the merger rate could potentially be higher than our upper limits. Conversely, for some parts of the parameter space, template-based searches have better sensitivities, although we expect them to lose sensitivity at higher eccentricities. Hence, including the VT from these searches \citep{LIGOScientific:2021usb, 2021arXiv211103606T} would tighten our upper limits. For simplicity, we limit our results to those from the cWB-eBBH analysis assuming all previously identified candidates are from quasicircular binaries.

Since binary mergers from dynamical formation channels can follow a mass distribution different from the one obtained from GWTC-3, we additionally computed $\mbox{VT}$ assuming other parameter distributions. We summarize our results in Table \ref{tab:population}. Our focus on high-mass, eccentric events can be particularly interesting for astrophysical formation channels that favor the production of both high mass and high eccentricity, such as gas-driven capture in AGN disks. For this scenario we adopted the AGN model of \cite{2021ApJ...920L..42G} as an illustrative example. Our search sensitivity for this model is marginally higher than for the GWTC-3 distribution because this model favors higher masses that are more likely to fall in the mass interval that we are most sensitive to in this analysis. Assuming non-detection of any eccentric event, we place a constraint of $<0.29$\,Gpc$^{-3}$ yr$^{-1}$ on the merger rate density at 90\% confidence level for AGN-assisted mergers. Taking an estimated $\sim 70\%$ of mergers being eccentric \citep{2022Natur.603..237S} and $\sim4\%$ of mergers having $M>70$\,$M_\odot$ \citep{2021ApJ...920L..42G}, we project the corresponding upper limit on the merger rate density to obtain upper limits on the overall AGN-assisted merger rate density as $\sim0.29$\,Gpc$^{-3}$~yr$^{-1}$\,$/ (0.7 \times 0.04) \sim 10.4$\,Gpc$^{-3}$ yr$^{-1}$. This is consistent with rate estimates in the literature (e.g., \citealt{2019ApJ...876..122Y,2021ApJ...920L..42G}). 

As a second illustrative model we used the distribution expected in dense star clusters (DSC), adopted from \cite{2021ApJ...921L..43Z}. For this population, we are able to place a constraint of $<0.34$\,Gpc$^{-3}$ yr$^{-1}$ on the merger rate density at 90\% confidence level assuming non-detection of any eccentric event. Taking an estimated $\sim 10\%$ being eccentric and $\sim18\%$ of mergers having $M>70$\,$M_\odot$,  we project the corresponding upper limit on the merger rate density to obtain upper limits on the overall DSC-assisted merger rate density as $\sim0.34$\,Gpc $^{-3}$~yr$^{-1}$\,$/ (0.1 \times 0.18) \sim 18.9$\,Gpc$^{-3}$ yr$^{-1}$. This is consistent with rate estimates in the literature \citep{Kremer:2019iul, 2021ApJ...921L..43Z}.

\begin{table}[ht]
\begin{center}
\begin{tabular}{ c  c  c  c }

\hline
 \multirow{2}*{$p(M)$} & \multirow{2}*{$p(q)$}  & \multirow{2}*{$p(e)$} & $\mbox{VT}$ \\
 & & & [Gpc$^{3}$yr] \\
\hline
 GWTC-3 & GWTC-3 & $2(1-e)$ & $6.88$ \\

 GWTC-3 & GWTC-3 & uniform & $6.93$ \\

 $M^{-2.3}$ & uniform & $2(1-e)$ & $8.22$ \\

 $M^{-2.3}$ & uniform & uniform & $8.27$ \\

 AGN & AGN & $2(1-e)$ & $7.85$ \\

 AGN & AGN & uniform & $7.91$ \\

 DSC & DSC & DSC & $6.69$ \\
\hline

\end{tabular}
\caption{Total volume--time covered by cWB-eBBH search assuming various source total mass, mass ratio, and eccentricity probability density functions for the different illustrative models described in Section \ref{sec:pop}.}
\label{tab:population}
\end{center}
\end{table}

\section{Conclusion}
\label{sec:conclusion}

We carried out a search that does not rely on template banks, and optimized it to be sensitive to high-mass ($M>70$\,$M_\odot$) eccentric binary black hole coalescences. We characterized the sensitivity for this search to understand our findings' implications for possible eccentric astrophysical populations. Our conclusions are as follows:
\begin{enumerate}\itemsep0em
\item We did not identify any high significance candidate that was not already detected by other searches. Our loudest and most significant new event has an IFAR of $1.32$~yr. We performed detailed follow-up for this event, and concluded that astrophysical origin could not be ruled out. However, our search results are consistent with the expected background for O3.
\item For our fiducial model, we adopted a mass distribution that assumes a Power Law + Peak model and best fits the observations listed in the GWTC-3 catalog. We also chose an eccentricity distribution (defined in Section \ref{sec:pop}) that favors quasi-circular binaries. For this assumed population, our search sensitivity is such that assuming non-detection of eccentric events, we can place a constraint of $<0.33$\,Gpc$^{-3}$ yr$^{-1}$ on the merger rate density at 90\% confidence level. 
This obtained overall sensitivity is similar to that of other searches for circular black hole mergers in a similar mass range (cf. inferred rate of $0.08^{+0.19}_{-0.07}$\,Gpc$^{-3}$ yr$^{-1}$ of mergers similar to GW190521; \citealt{O3_IMBH}). 
\item  
As an illustrative example, we found that non-detection of any eccentric event corresponds a constraint of $< 10.4$\,Gpc$^{-3}$ yr$^{-1}$ on the AGN-assisted merger rate density, consistent with rate estimates in the literature (e.g., \citealt{2019ApJ...876..122Y,2021ApJ...920L..42G}). 
\item As a second illustrative model, we computed our search sensitivity to mergers in dense star clusters, considering the model of \cite{2021ApJ...921L..43Z}. The results are similar to the AGN channel and our expected sensitivity for a generic eccentric model. For this model, we found that non-detection of eccentric events corresponds to a constraint of $<18.9$\,Gpc$^{-3}$ yr$^{-1}$ on the merger rate density, consistent with rate estimates in literature \citep{Kremer:2019iul, 2021ApJ...921L..43Z}.
\end{enumerate}

The constraints we place on the rate of eccentric binary coalescences in this work are significantly improved over those computed with data obtained from the first and second observing runs \citep{Abbott_2019_ebbh}. This improvement can be attributed to increased sensitivity of the detectors, progress in the development of highly accurate eccentric waveforms in the high mass domain, and an optimized eccentric search. In view of the expected sensitivity of the fourth observing run by LIGO--Virgo--KAGRA \citep{KAGRA:2013rdx}, we anticipate to see a significant rise in the number of binary black hole detections. This increases our prospects of detecting gravitational-wave signals from eccentric binary coalescences. Regardless, a non-detection would enable us to further constrain the binary black hole merger rates in astrophysical models favouring eccentric orbits. 

Future works will need to expand the study to  eccentricities greater than 0.3, and to include masses below $70\,M_\odot$ as well as black hole spins.


\acknowledgements
Data-quality products and event-validation results were computed using the \DQR{}~\cite{DQRdocumentation}, \DMT{}~\cite{DMTdocumentation},  \GWDETCHAR{}~\cite{gwdetchar-software}, \HVETO{}~\cite{Smith:2011an} and \IDQ{}~\cite{Essick:2020qpo} software packages and contributing software tools.  
Analyses in this paper relied upon the \LALSUITE{} software library~\cite{lalsuite-software}.
The detection of the signals and subsequent significance evaluations in this paper were performed with the \ac{CWB}~\cite{Klimenko:2005xv,cWB_MaxL} package. 
Estimates of the noise spectra and glitch models were obtained using \BAYESWAVE{}~\cite{Cornish:2014kda,Littenberg:2015kpb,Cornish:2020dwh}. 
Source-parameter estimation was performed with the \LALINFERENCE{}~\cite{2015PhRvD..91d2003V} library.
\PESUMMARY{} was used to post-process and collate parameter-estimation results~\cite{Hoy:2020vys}.
Plots were prepared with \PLT{}~\cite{Hunter:2007ouj} and \GWPY{}~\cite{gwpy-software}. 
\NUMPY{}~\cite{Harris:2020xlr} and \SCIPY{}~\cite{Virtanen:2019joe} were used in the preparation of the manuscript.

%
\newif\ifcoreonly\coreonlytrue
\newif\ifkagra\kagratrue
\newif\ifheader\headerfalse
This material is based upon work supported by NSF’s LIGO Laboratory which is a major facility
fully funded by the National Science Foundation.
The authors also gratefully acknowledge the support of
the Science and Technology Facilities Council (STFC) of the
United Kingdom, the Max-Planck-Society (MPS), and the State of
Niedersachsen/Germany for support of the construction of Advanced LIGO 
and construction and operation of the GEO\,600 detector. 
Additional support for Advanced LIGO was provided by the Australian Research Council.
The authors gratefully acknowledge the Italian Istituto Nazionale di Fisica Nucleare (INFN),  
the French Centre National de la Recherche Scientifique (CNRS) and
the Netherlands Organization for Scientific Research (NWO), 
for the construction and operation of the Virgo detector
and the creation and support  of the EGO consortium. 
The authors also gratefully acknowledge research support from these agencies as well as by 
the Council of Scientific and Industrial Research of India, 
the Department of Science and Technology, India,
the Science \& Engineering Research Board (SERB), India,
the Ministry of Human Resource Development, India,
the Spanish Agencia Estatal de Investigaci\'on (AEI),
the Spanish Ministerio de Ciencia e Innovaci\'on and Ministerio de Universidades,
the Conselleria de Fons Europeus, Universitat i Cultura and the Direcci\'o General de Pol\'{\i}tica Universitaria i Recerca del Govern de les Illes Balears,
the Conselleria d'Innovaci\'o, Universitats, Ci\`encia i Societat Digital de la Generalitat Valenciana and
the CERCA Programme Generalitat de Catalunya, Spain,
the National Science Centre of Poland and the European Union – European Regional Development Fund; Foundation for Polish Science (FNP),
the Swiss National Science Foundation (SNSF),
the Russian Foundation for Basic Research, 
the Russian Science Foundation,
the European Commission,
the European Social Funds (ESF),
the European Regional Development Funds (ERDF),
the Royal Society, 
the Scottish Funding Council, 
the Scottish Universities Physics Alliance, 
the Hungarian Scientific Research Fund (OTKA),
the French Lyon Institute of Origins (LIO),
the Belgian Fonds de la Recherche Scientifique (FRS-FNRS), 
Actions de Recherche Concertées (ARC) and
Fonds Wetenschappelijk Onderzoek – Vlaanderen (FWO), Belgium,
the Paris \^{I}le-de-France Region, 
the National Research, Development and Innovation Office Hungary (NKFIH), 
the National Research Foundation of Korea,
the Natural Science and Engineering Research Council Canada,
Canadian Foundation for Innovation (CFI),
the Brazilian Ministry of Science, Technology, and Innovations,
the International Center for Theoretical Physics South American Institute for Fundamental Research (ICTP-SAIFR), 
the Research Grants Council of Hong Kong,
the National Natural Science Foundation of China (NSFC),
the Leverhulme Trust, 
the Research Corporation,
the National Science and Technology Council (NSTC), Taiwan,
the United States Department of Energy,
and
the Kavli Foundation.
The authors gratefully acknowledge the support of the NSF, STFC, INFN and CNRS for provision of computational resources.

This work was supported by MEXT, JSPS Leading-edge Research Infrastructure Program, JSPS Grant-in-Aid for Specially Promoted Research 26000005, JSPS Grant-inAid for Scientific Research on Innovative Areas 2905: JP17H06358, JP17H06361 and JP17H06364, JSPS Core-to-Core Program A. Advanced Research Networks, JSPS Grantin-Aid for Scientific Research (S) 17H06133 and 20H05639 , JSPS Grant-in-Aid for Transformative Research Areas (A) 20A203: JP20H05854, the joint research program of the Institute for Cosmic Ray Research, University of Tokyo, National Research Foundation (NRF), Computing Infrastructure Project of Global Science experimental Data hub Center (GSDC) at KISTI, Korea Astronomy and Space Science Institute (KASI), and Ministry of Science and ICT (MSIT) in Korea, Academia Sinica (AS), AS Grid Center (ASGC) and the National Science and Technology Council (NSTC) in Taiwan under grants including the Rising Star Program and Science Vanguard Research Program, Advanced Technology Center (ATC) of NAOJ, and Mechanical Engineering Center of KEK.

{\it We would like to thank all of the essential workers who put their health at risk during the COVID-19 pandemic, without whom we would not have been able to complete this work.}

\appendix
\section{Post-Production Vetoes}
\label{sec:vetoes}

In this appendix, we will describe in detail the post-production vetoes that are applied by the standard cWB pipeline \citep{gayathri_wavegraph, Lopez:2021ikt} to distinguish between true gravitational-wave signals and non-Gaussian noise artifacts that can mimic gravitational-wave signals.

The first set of vetoes are based on the morphology of the reconstructed signals. These vetoes are applied to the following cWB summary statistics: the energy-weighted central frequency of the signal $f_0$; $\mathcal{M^*}$ which is the reconstructed chirp mass parameter is obtained by fitting the signal with the characteristic time--frequency evolution for a quasi-circular binary ($f\propto (t-t_{c})^{-3/8}$), and $Q_a$, the waveform shape parameter introduced in Section \ref{sec:optimization}. $Q_a$ is a function of the cWB parameter Qveto \citep{Qveto, gayathri_wavegraph, Mishra:2021tmu}, which quantifies how well the total energy of the signal is distributed across time. The first set of vetoes removes events that do not satisfy $24$\,Hz $<f_0 < 100$\,Hz, $\abs{\mathcal{M^*}/\mathrm{M_{\odot}}}>10$, $\abs{(\mathcal{M^*}/\mathrm{M_{\odot}})/Q_{a}^2}>15$, $\mathcal{M^*/\mathrm{M_{\odot}}}>-100$.

The next set of vetoes are based on cWB reconstruction, and the correlation of the event across the network of detectors. The cWB summary statistics involved in this set are: norm, defined as the ratio between the total energy over all wavelet resolution levels used for the analysis and the reconstructed energy of the event; $\chi^2$, a parameter that quantifies the quality of signal reconstruction by computing the residual noise energy that remains once the reconstructed signal is subtracted from data \citep{gayathri_wavegraph}, and finally the $c_{c}[0]$ and $c_{c}[2]$ parameters that describe the correlation of the signal across the network of detectors in time domain and frequency domain, respectively \citep{Tiwari_2015}. The second set of vetoes remove candidate events that do not satisfy $\mathrm{norm}>4$, $\log_{10}(\chi^2)<0.4$, $c_{c}[0]>0.8$, $c_{c}[2]>0.7$.

The two sets of vetoes described above were optimized with gravitational waveforms for quasi-circular binary black hole coalescences for the standard cWB pipeline. We found that they performed optimally in recovering eBBH signals as well. Therefore, these vetoes along with the new eBBH veto introduced in Section \ref{sec:optimization} were chosen as the final set of vetoes for the cWB-eBBH search pipeline.

\bibliography{Refs}
\end{document}